\documentclass[twocolumn,epjc3]{svjour3}
\usepackage[T1]{fontenc}
\usepackage{amsmath,amssymb}
\usepackage{newtxtext,newtxmath}
\usepackage{graphicx}
\usepackage[numbers]{natbib}
\usepackage{caption}
\usepackage{microtype}
\usepackage{hyphenat}
\usepackage[colorlinks,citecolor=blue,urlcolor=blue,linkcolor=blue]{hyperref}

\smartqed  

\journalname{Eur. Phys. J. C}

\begin{document}
	
	\title{Absorption and scattering of massless scalar waves by black holes in quasi-topological gravity}
	\author{Sihao Fan\thanksref{addr1} 
		\and 
		Chen Wu\thanksref{e1,addr2}
		\and
		Wenjun Guo\thanksref{addr1}
	}
	
	\thankstext{e1}{e-mail: wuchenoffd@gmail.com}

	\institute{University of Shanghai for Science and Technology, Shanghai 200093, China \label{addr1} 
		\and
		Xingzhi College, Zhejiang Normal University, Jinhua 321004, Zhejiang, China \label{addr2}}
	\date{Received: date / Accepted: date}
	
	\maketitle
	
	\begin{abstract}
		
		{We consider massless scalar field perturbations of regular black holes in quasi-topological gravity.After reviewing various regular black hole solutions, we compute the total absorption cross-section and the partial absorption cross-section for different choices of the parameters using the partial wave method.The results show that the larger the parameter $\alpha$, introduced in the context of quasi-topological gravity, the smaller the total absorption cross-section.In addition, we find that the influence of the space-time dimension $D$ on the absorption and scattering of massless scalar waves in this set of regular black holes differs from an earlier computations.}
		
	\end{abstract}
	
	\section{Introduction} 
	Black holes are among the most significant objects of study in cosmology and represent one of the hallmark predictions of general relativity. In recent years, significant progress has been made in black hole observations. For example , the  Event Horizon Telescope (EHT) successfully captured the shadow image of the supermassive black hole at the center of galaxy M87 \cite{EHT2019VI,EHT2019I}, and the LIGO-VIRGO collaboration detected gravitational wave signals generated during the merger of binary black holes, successfully verifying general relativity's predictions for spacetime dynamics in strong gravitational fields \cite{Abbott2016,Abbott2017}. These observations confirmed the existence of black holes and also highlighted the importance of general relativity in black hole research. Furthermore, these discoveries have stimulated in-depth research interest in black hole absorption and scattering processes. Currently, there are numerous related studies, such as the absorption and scattering characteristics of massless scalar fields in black hole backgrounds \cite{Liao2012,Vieira2016,DeOliveira2018a}, scattering and absorption in higher dimensional black holes \cite{Chen2011,Jorge2015,Crispino2010}, and related research on charged black holes \cite{Benone2014,Benone2017,DeOliveira2018b}. On the theoretical level,the absorption and scattering properties of massless scalar fields, higher-dimensional spacetime, and charged black holes have been extensively explored, but systematic research on the scattering dynamics of novel black hole models is still lacking. This study focuses on a recently proposed new black hole model and are solutions of Quasi-topological gravities \cite{Bueno2025}. We employ the partial wave method to calculate its scattering and absorption cross-sections, and examine how the model parameters affect the results.
	
	  This paper will further investigate the quasi-normal  modes (QNMs) of the studied black hole, which provide insights into  the interaction between the black hole and surrounding fields. QNMs are determined by the eigenvalues of perturbation equations subject to  specific boundary conditions and dominate the field decay behavior in black hole backgrounds. As important macroscopic characteristics of black holes, QNMs play a crucial role in gravitational wave research and have become a hot topic in recent years. Studies show that QNMs dominate the exponentially decaying ringdown phase in gravitational wave signals from perturbed black holes \cite{Kokkotas1999}, and govern the ringdown signals in gravitational waves from newly formed black holes after binary mergers \cite{Pretorius2005,Campanelli2006} . In 2015, the Laser Interferometer Gravitational Wave Observatory (LIGO) first detected gravitational waves produced by the merger of a binary black hole system \cite{Abbott2016,Abbott2017}, a discovery considered as the dawn of gravitational wave astronomy. These significant advancements have stimulated extensive research on QNMs under various spacetime and physical field backgrounds. This paper employs the WKB approximation method to calculate the QNMs of the regular black hole modified by the quasi-normal renormalization group  method, and present  corresponding numerical results and analysis.
	
	The formulation of Hawking radiation theory initiated of black hole thermodynamics and quantum theoretical research. Remarkably, the represents a profound synthesis of Hawking radiation synthesizes fundamental principles from general relativity, quantum mechanics, and thermodynamics. The theoretical framework for computing Hawking radiation was first established by Damour and Ruffini in 1976 \cite{Damour1976}. Subsequent developments included Parikh and Wilczek's quantum tunneling approach in 2000 \cite{Parikh2000}, which was subsequently refined by Kerner and Mann \cite{Kerner2008}. In 2011, Kokkotas et al. calculated Hawking radiation through greybody factors based on the sixth-order Wentzel-Kramers-Brillouin (WKB) method proposed by Konoplya \cite{Konoplya2003,Kokkotas2011}. This methodology continues to serve as a fundamental tool for investigating black hole absorption processes \cite{Sun2023}.
	
	General relativity and quantum mechanics describe gravitational interactions and atomic-scale physical laws respectively, yet their theoretical frameworks remain fundamentally incompatible. Since Hawking's seminal discovery that black holes can absorb particles, the absorption cross-section of black holes has emerged as a key research focus. Sanchez established that the absorption cross-section of massless  scalar fields exhibits oscillatory behavior near the geometric optics limit \cite{Sanchez1978}. Unruh investigated non-rotating black holes and found the Dirac particle absorption cross-section to be approximately 1/863 of that for massless scalar particles \cite{Unruh1976}. Moreover, as previously mentioned, extensive studies have explored interactions between various field types and black holes through modified field equations and metrics, suggesting an urgent need for in-depth analysis of regulatory mechanisms governing absorption processes through metric characteristics and field equation formulations.
	
    Higher-dimensional black holes have emerged as a pivotal component in quantum gravity research. Identifying the distinctive signatures of black holes in extra dimensions is a critical endeavor, achievable through a comprehensive analysis of astrophysical observations. To this end, several sophisticated methods have been developed, including explorations of black hole shadows \cite{papnoi2014,singh2017}, investigations of gravitational lensing and optical properties \cite{abdujabbarov2015,belhaj2020}, examinations of quasinormal modes \cite{matyjasek2021,yan2020}, studies of thermodynamic behavior \cite{andre2021,barman2020,chen2008}, and analyses of other related phenomena \cite{ahmedov2021,ishibashi2003,kunstatter2014,lake2003}. Scattering studies show particular promise for unveiling extra-dimensional gravitational effects. Recent advances in laboratory analogue black holes \cite{Dolan2009,Crispino2007} offer experimental validation platforms. Crucially, these setups demand high-precision computations of scattering cross-sections and quasinormal spectra to guide experimental design and interpret observational data.

	This work bridges these gaps by quantifying scattering / absorption cross-sections and QNM spectra in regular black holes via partial wave analysis and WKB methods. We assess parameter modification impacts on dynamics while probing potential dark matter and gravitational wave connections.
	
	\section{Wave Analysis Methodology}
	\subsection{Regular black holes from pure gravity}
	In this section, we will first introduce the regular black holes used in this study. The line element for the D-dimensional spherically symmetric regular black hole spacetime solution is given as follows:
	\begin{equation}{\label{eq1}}
		ds^2=-f(r) c^2 dt^2+f^{-1}(r)dr^2+r^2 d\Omega_{(D-2)}^2,
	\end{equation}

	When $D=4$,$d\Omega_{(D-2)}^2=d\theta^2+sin^2 \theta d\psi^2$(two dimensional sphere), and the line element reduces to that of a four dimensional spherically symmetric regular black hole. Here, $f(r)$ is the metric function. Critically, the choice of metric function governs the energy-momentum tensor construction, which in turn determines the spacetime geometry.
	
	This class of black holes emerges from foundational investigations of spacetime singularities \cite{Bueno2025}. Specifically, Bueno et al. demonstrated that within quasitopological gravity with finite-order curvature corrections, the divergence strength of black hole singularities can be substantially suppressed. Their analysis revealed that when increasingly higher-order relativistic gravitational couplings satisfy certain constraints, the resulting solutions admit regular (singularity-free) black holes. We now reconstruct their derivation procedure.
	
	{The action of a Quasi-topological theory is
		\begin{equation}
			I_{\mathrm{QT}} = \frac{1}{16\pi G} \int \mathrm{d}^{D} x \sqrt{|g|} \left[ R + \sum_{n=2}^{n_{\max}} \alpha_{n} \mathcal{Z}_{n} \right],
		\end{equation}
		Where $\alpha_{n}$ are arbitrary couping constants with dimension of length$^{2(n-1)}$.
		The equations of motion of \eqref{eq1} imply that
		\begin{equation}
			\frac{d}{dr}[r^{D-1} h(\psi)]=0,
		\end{equation}
		Where
		
		\begin{equation}{\label{eq_add4}}
			h(\psi) \equiv \psi + \sum_{n=2}^{n_{\max}} \alpha_{n} \psi^{n}, \quad \psi \equiv \frac{1 - f(r)}{r^{2}},
		\end{equation}
		For $n_{max}= N $ there is a curvature singularity at $r=0$ where Kretschmann scalar diverges.However the rate of divergence decreases as $N$ increases.This behavior motivates considering  an infinite tower of higher-curvature corrections by taking the upper summation limit $n_{\text{max}} \rightarrow \infty$ in \eqref{eq_add4}.
		
		Table~\ref{tab1} compiles the metric functions given by Bueno \cite{Bueno2025} for distinct quasi-topological couplings.
		The models employed in this study include:}

\begin{table}[htbp]
	\caption{\label{tab1}
		The metric functions exhibit distinct behaviors under different values of quasi-topological couplings $\alpha_{n}$.
	}
	\renewcommand{\arraystretch}{2}
	\setlength{\tabcolsep}{14pt} 
	\begin{tabular}{l l}  
		\hline
		 $\alpha_{n}$ & $f(r)$ \\ 
		\hline
		$\alpha^{n-1}$ & $1-\frac{mr^2}{ r^{D-1}+\alpha m}$  \\  
		$\frac{\alpha^{n-1}}{n}$ & $1-\frac{r^2}{\alpha}  (1-e^{-\alpha m/r^{D-1}}) $  \\  
		$n\alpha^{n-1}$ & $1-\frac{2mr^2}{r^{D-1}+2\alpha m+\sqrt{ r^{2(D-1)} +4\alpha mr^{D-1} }}$ \\  
		$\frac{(1-(-1)^{n})}{2}\alpha^{n-1}$ & $1-\frac{2mr^2}{r^{D-1}+\sqrt{r^{2(D-1)} +4\alpha ^2 m^2 }}$ \\  
		$\dfrac{\left(1 - (-1)^{n}\right) \Gamma\left(\frac{n}{2}\right)}
		{2\sqrt{\pi}\,\Gamma\left(\frac{n + 1}{2}\right)} \alpha^{n - 1}
		$ & $1-\dfrac{mr^2}{\sqrt{r^{2(D-1)} +\alpha ^2 m^2}}$ \\  
		\hline
	\end{tabular}
\end{table}
{%
	\begin{equation}{\label{eq2}}
		f(r)=1-\frac{(mr^2)}{ (r^{D-1}+\alpha m)}  ,
	\end{equation}
	\begin{equation}{\label{eq3}}
	f(r)=1-\frac{r^2}{\alpha}  (1-e^{-\alpha m/r^{D-1}) },
    \end{equation}
	\begin{equation}{\label{eq4}}
	f(r)=1-\frac{2mr^2}{r^{D-1}+2\alpha m+\sqrt{ r^{2(D-1)} +4\alpha mr^{D-1} }} ,
    \end{equation}
	\begin{equation}
	f(r)=1-\frac{2mr^2}{r^{D-1}+\sqrt{r^{2(D-1)} +4\alpha ^2 m^2 }} ,
    \end{equation}
	\begin{equation}
	f(r)=1-\frac{mr^2}{\sqrt{r^{2(D-1)} +\alpha ^2 m^2}}  ,
	\end{equation} 
	
    {Where m is an integration constant which is proportional to the Arnowitt-Deser-Misner(ADM) mass of the solution.} These metrics contain two parameters: the dimensionality D and the coupling constant $\alpha$ which characterizes the higher-order curvature corrections. The magnitude of $\alpha$ governs the contribution weight of these correction terms to the gravitational field equations. Larger $\alpha$ values amplify the significance of higher-order curvature effects, whereas smaller $\alpha$ values correspond to diminished influence. Crucially, this framework achieves singularity resolution in black holes by incorporating an infinite sequence of higher-order curvature correction terms (formally through the limit n→$\infty$), thereby ensuring spacetime regularity within the black hole interior.
	
	It can be observed that for Eq.~({\ref{eq2}}), the case D=4 recovers the Hayward black hole solution \cite{Hayward2006}. In this study, we extend the solution to higher dimensions $D \geq 5$, where the metric incorporates an infinite tower of higher-curvature terms, constituting exact solutions in pure gravitational theory. Regarding Eq.~({\ref{eq3}}), its functional form exhibits similarities with the Dymnikova black hole solution \cite{Dymnikova2004}, as both employ exponential suppression to regularize singularities. However, our framework distinguishes itself by providing explicit metrics for $D \geq 5$ spacetime dimensions.
	
	Notably diverging from prior investigations, these black hole solutions inherently satisfy the first law of black hole thermodynamics \cite{Abramowitz1964}.Furthermore, their adherence to Birkh-off's theorem \cite{Schutz1985} significantly simplifies stability analysis through linear perturbation techniques. A critical theoretical nuance arises from the absence of consistent quasi-topological gravitational theories in $D=4$ spacetime dimensions. Nevertheless, recent foundational work by Bueno et al. \cite{Bueno2025} has established a well-defined prescription for dimensional reduction, enabling the physical viability of studying $D=4$ solutions as limiting cases. Our analysis will explicitly demonstrate this framework's predictive power by presenting new numerical results derived from the $D=4$ configuration.
	
	\begin{figure}
		\centering\hfil\includegraphics[scale=0.18]{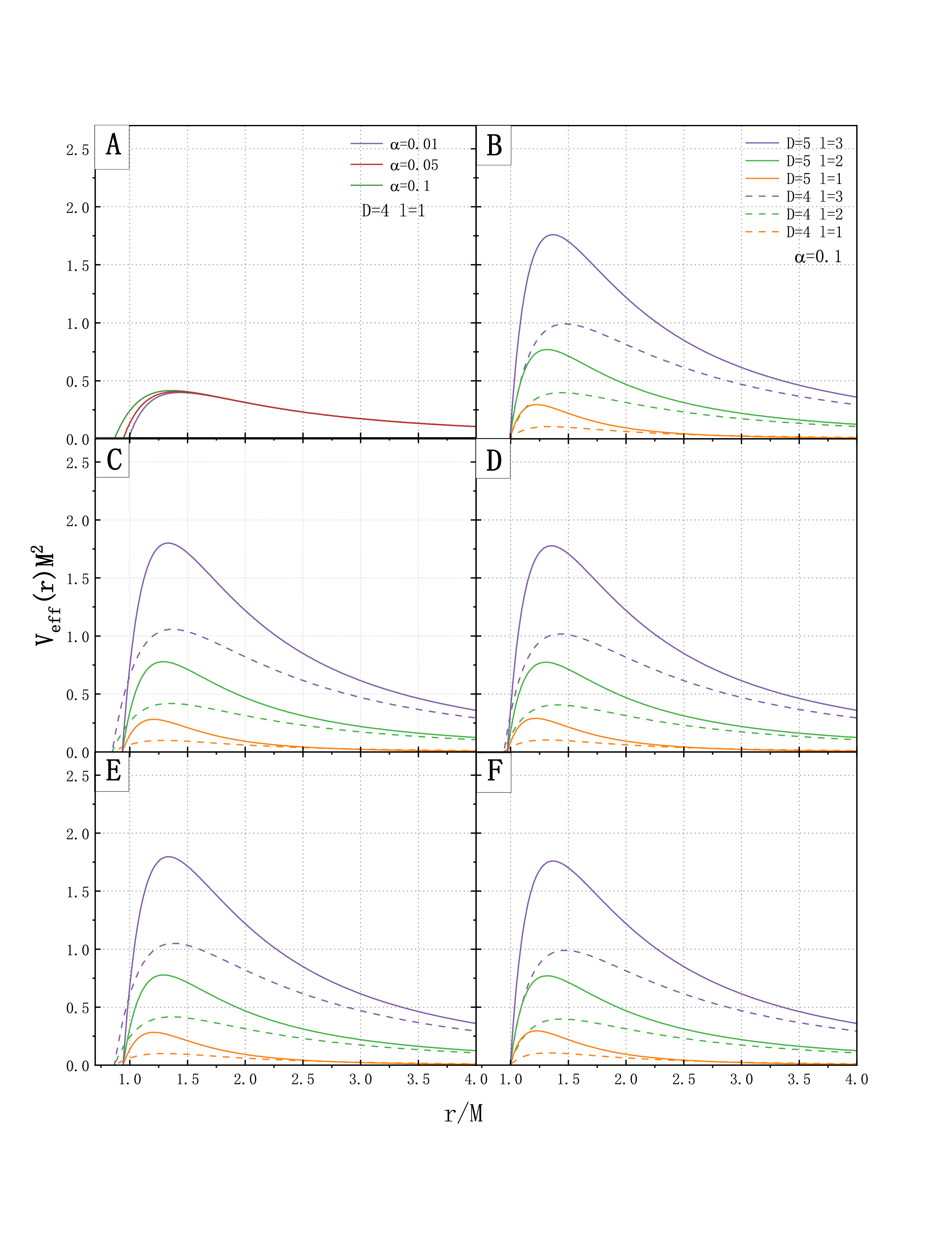}
		\caption{\label{fig1}The figure presents the effective potential profiles: Panel (A) corresponds to Eq.~({\ref{eq2}}) with parameter configurations . Panels (B)-(F) display effective potentials derived from the five distinct metrics discussed earlier.}
	\end{figure}
	
	\begin{figure}
	\noindent\hfil\includegraphics[scale=0.18]{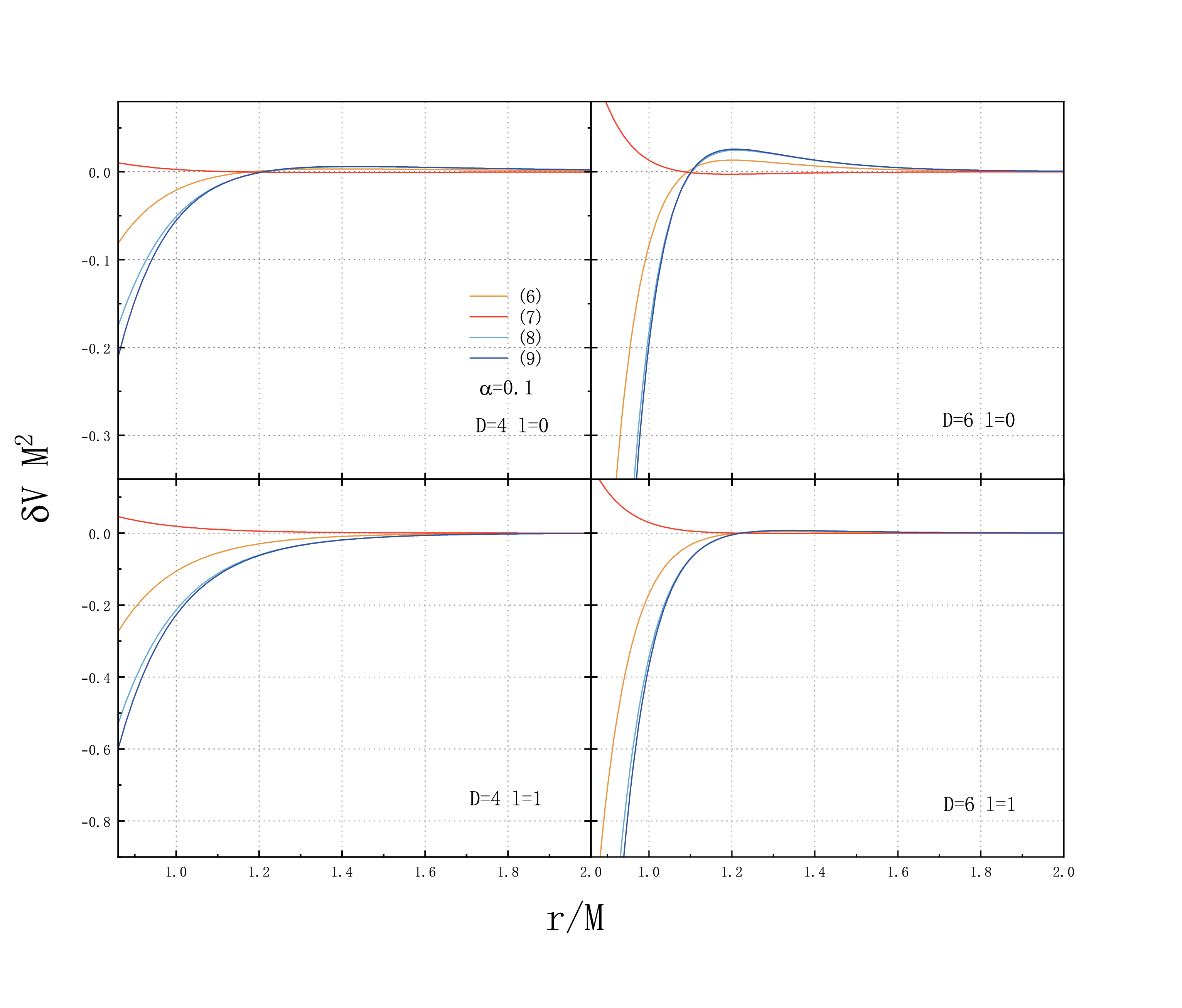}
	\caption{\label{fig2}The figure illustrates the effective potential differences across varying spacetime dimensions $D$ and angular momentum quantum numbers $l$, with respect to Eq.~({\ref{eq2}}) as the reference baseline.}
    \end{figure}

	Fig.~{\ref{fig1}} illustrates the effective potential in spherical coordinates under different angular momenta and parameters. It can be observed that the effective potential vanishes at the black hole horizon and asymptotically approaches zero as $ r \to \infty $. Additionally, it is found that the potential barrier becomes higher with increasing angular quantum number $ l $, while an increase in the spatial dimension $ D $ also leads to a higher potential barrier . Fig.~{\ref{fig1}} A shows that parameter $\alpha$ has a small effect on the effective potential energy, and it is almost undetectable at large r . From Figs.~{\ref{fig1}} C,D,and E, it is evident that the dimensional parameter $ D $ has a significant impact on the black hole horizon. As $ D $ increases, the size of the black hole horizon also increases, though the magnitude of this effect gradually diminishes with higher values of $ D $. However, intriguingly, in Figs.~{\ref{fig1}} B and F, the dimensional parameter $ D $ does not exhibit a noticeable influence on the size of the black hole horizon.
	
	Fig.~{\ref{fig2}} demonstrates the effective potential difference between distinct metrics under varying spacetime dimensions $ D $ and angular momentum quantum numbers $ l $. It can be observed that at larger $ r $, the potential difference asymptotically approaches zero, which aligns with the trend in Fig.~{\ref{fig1}} where the effective potentials of different metrics exhibit diminishing discrepancies as $ r $ moves away from the black hole horizon. Furthermore, the region of significant potential difference is predominantly localized in the vicinity of the black hole horizon.
	
	Furthermore, it is observed that the effective potential in Eq.~({\ref{eq3}}) exhibits the closest proximity to that in Eq.~({\ref{eq2}}), leading to the inference that their scattering and absorption cross-sections share the most analogous behavior. This conclusion is further corroborated by the results displayed in Fig.~{\ref{fig4},\ref{fig5}}. From the figure, one can discern that the disparity in effective potentials correspondingly amplifies with increasing values of the spacetime dimension $ D $ and angular quantum number $ l $.
	
	\subsection{partial wave method}
	This work investigates the absorption and scattering characteristics of massless scalar waves in a black hole background, governed by the Klein-Gordon equation:
	\begin{equation}{\label{eq7}}
		\frac{1}{\sqrt[]{-g}}  \partial_\mu  (\sqrt[]{-g} g^{\mu\nu } \partial_\nu \Psi)=0,
	\end{equation}
	The solution to Eq.~(\ref{eq7}) can be decomposed as:
	\begin{equation}
		\Phi = \frac{\Psi_{\omega l}}{r} Y_{l,m}(\theta, \phi) e^{-i \omega t},
	\end{equation}
	where the angular part $ Y_{l m}(\theta, \phi) $ represents the scalar spherical harmonics with $ l $ and $ m $ being the angular momentum quantum numbers. Our primary focus lies in the radial part, which satisfies the differential equation:
	
	\begin{equation}{\label{eq9}}
		\frac{1}{f} \frac{\partial^2 \Psi}{\partial t^2} 
		- \frac{1}{r^2} \frac{\partial}{\partial r} \left( f r^2 \frac{\partial \Psi}{\partial r} \right) 
		+ \frac{1}{r^2} \nabla^2 \Psi = 0,
	\end{equation}
    Eq.~(\ref{eq9}) can be further simplified as:
    \begin{equation}{\label{eq10}}
	f \frac{d}{dr} \left( f \frac{d\Psi_{\omega l}}{dr} \right) + \left[ \omega^2 - V_{\text{eff}}^l(r) \right] \Psi_{\omega l} = 0,
    \end{equation}
    We may introduce the tortoise coordinate $ \chi $, defined by:
    \begin{equation}
	\chi = \int f^{-1}(r) \, dr,
    \end{equation}
    which transforms Eq.~(\ref{eq10}) into:
    \begin{equation}{\label{eq12}}
  	    \frac{d^2 \Psi_{\omega l}}{d\chi^2} + \left[ \omega^2 - V_{\text{eff}}^l(\chi) \right] \Psi_{\omega l} = 0,
    \end{equation}
    where the effective potential is given by:
\begin{equation}
	V_{\text{eff}}^l(\chi) = f \left[ \frac{f'}{r} + \frac{l(l+1)}{r^2} \right].
\end{equation}

  In Fig.~{\ref{fig1}}, we plot the functional dependence of $ V_{\text{eff}} $ on the coordinate $ r $. It can be observed that as the angular quantum number $ l $ increases, the peak value of the scattering potential progressively rises. In canonical black hole spacetimes with distinct metrics, fixing $ l $ while increasing the spacetime dimension $ D $ similarly leads to an enhancement of the scattering potential's peak. Furthermore, the figure reveals that the discrepancies in potential values under varying parameters gradually diminish with increasing $ r $.
  
  It is evident that Eq.~(\ref{eq12}) bears a striking resemblance to the Schr\"{o}dinger equation in quantum mechanics. In fact, when studying the scattering of matter fields in black hole spacetimes, we adopt methods analogous to those used for matter wave scattering in quantum mechanics. Based on the characteristics of the effective scattering potential, the solution to this equation can be expressed as:

\begin{equation}\label{eq14}
	\footnotesize
	\Psi_{\omega l} \approx 
	\begin{cases}
		T_{\omega l}(\omega) \, e^{-i\omega\chi} & \chi \to -\infty; \\
		\omega\chi \left[ (-i)^{l+1} h_l^{(1)*}(\omega\chi) + (i)^{l+1} R_{\omega l}(\omega) h_l^{(1)}(\omega\chi) \right] & \chi \to +\infty;
	\end{cases}
\end{equation}

where $ h_l^{(1)}(\omega\chi) $ represents the spherical Bessel function of the third kind . The quantities $ |T_{\omega l}|^2 $ and $ |R_{\omega l}|^2 $ correspond to the transmission coefficient and reflection coefficient, respectively, satisfying the probability conservation equation:

\begin{equation}
	|T_{\omega l}|^2 + |R_{\omega l}|^2 = 1,
\end{equation}
    Considering the asymptotic form of the spherical Hankel function of the first kind, when $ |\omega\chi| \gg \frac{l(l+1)}{2} $, the following relation holds:
     \begin{equation}
     	h_l^{(1)}(\omega\chi) = j_l(\omega\chi) + i n_l(\omega\chi) \approx (-i)^{l+1} \frac{e^{i\omega\chi}}{\omega\chi},
     \end{equation}
     Therefore, Eq.~({\ref{eq14}}) can be approximated as:
     \begin{equation}
     	\Psi_{\omega l} \approx 
     	\begin{cases}
     		A_{\omega l}^{\text{tr}}(\omega) \, e^{-i\omega\chi} & \chi \to -\infty; \\
     		A_{\omega l}^{\text{in}} \, e^{-i\omega\chi} + A_{\omega l}^{\text{out}} \, e^{+i\omega\chi} & \chi \to +\infty;
     	\end{cases}
     \end{equation}
     The coefficients $ A_{\omega l}^{\text{tr}}(\omega) $, $ A_{\omega l}^{\text{in}}(\omega) $, and $ A_{\omega l}^{\text{out}}(\omega) $ are related to the incident, reflected, and absorbed components of the partial waves. The phase shift$ \delta_l $ can be defined as:
     \begin{equation}
     	e^{2i\delta_l} = (-1)^{l+1} \frac{A_{\omega l}^{\text{out}}}{A_{\omega l}^{\text{in}}},
     \end{equation}
     Typically, we normalize the transmission coefficient such that $ |A_{\omega l}^{\text{tr}}|^2 = 1 $, leading to:
     \begin{equation}
     	1 + |A_{\omega l}^{\text{out}}|^2 = |A_{\omega l}^{\text{in}}|^2.
     \end{equation}
     The relationship between these coefficients and the transmission/reflection amplitudes is:
     \begin{equation}
     	T_{\omega l} = \frac{1}{A_{\omega l}^{\text{in}}}, \quad R_{\omega l} = \frac{A_{\omega l}^{\text{out}}}{A_{\omega l}^{\text{in}}}.
     \end{equation}
     \subsection{Quasi-normal modes}
     QNMs describe the characteristic vibrational modes of a black hole under external perturbations. In the WKB approximation \cite{Schutz1985}, they are given by:
     \begin{equation}
     	\omega^2 \approx V_{\text{eff}}^l(r_0) - i \left(n + \frac{1}{2}\right) \sqrt{-2 V_{\text{eff}}''(r_0)},
     \end{equation}
     where $ r_0 $ is the extremum of the potential barrier, and $ n $ is the overtone number.

     \begin{figure}
     	\noindent\hfil\includegraphics[scale=0.18]{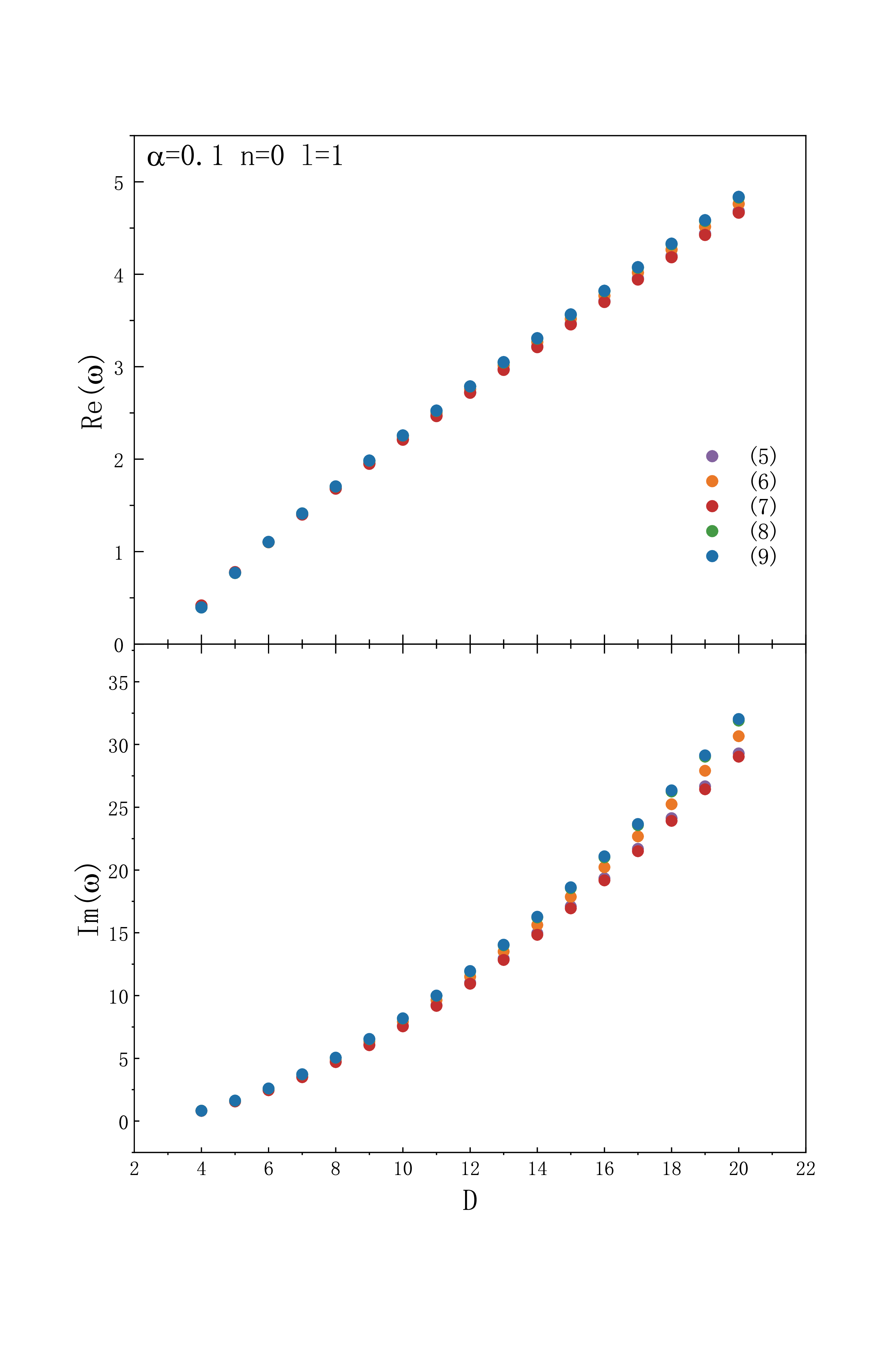}
     	\caption{\label{fig3}The figure above presents the numerical results for the real part $ \text{Re}(\omega) $ and imaginary part $ \text{Im}(\omega) $ of the QNMs corresponding to the five aforementioned metrics.}
     \end{figure}
     
     Fig.~\ref{fig3} demonstrates that as the spacetime dimension $D$ increases, both the real part Re($\omega$) and imaginary part Im($\omega$) of the QNMs show significant enhancement, revealing the decisive influence of dimension $D$ on the spectral characteristics of QNMs. This phenomenon suggests that higher-dimensional spacetime amplifies the field's damping effects (dominated by the imaginary part Im($\omega$)). Further analysis indicates that increasing dimension $D$ also leads to a marked expansion in the QNM differences between distinct black hole metrics, which implies that studying higher-dimensional QNMs can provide crucial criteria for distinguishing between different black hole solutions and holds significant value for verifying theoretical predictions in higher-dimensional gravitational theories.
     \section{Numerical Results and Analysis}
     \subsection{Absorption Cross-section}
     In quantum mechanics, the total absorption cross-section can be expressed as:
     \begin{equation}
     	\sigma_{\text{abs}} = \sum_{l=0}^{\infty} \sigma_{\text{abs}}^{(l)},
     \end{equation}
     where $\sigma_{\text{abs}}^{(l)}$ is the partial absorption cross-section:
     \begin{equation}
     	\sigma_{\text{abs}}^{(l)} = \frac{\pi}{\omega^2} (2l + 1) \left(1 - \left| e^{2i\delta_l} \right|^2 \right) = \frac{\pi}{\omega^2} (2l + 1) \left| T_{\omega l} \right|^2.
     \end{equation}
     
     \begin{figure}
     	\noindent\hfil\includegraphics[scale=0.19]{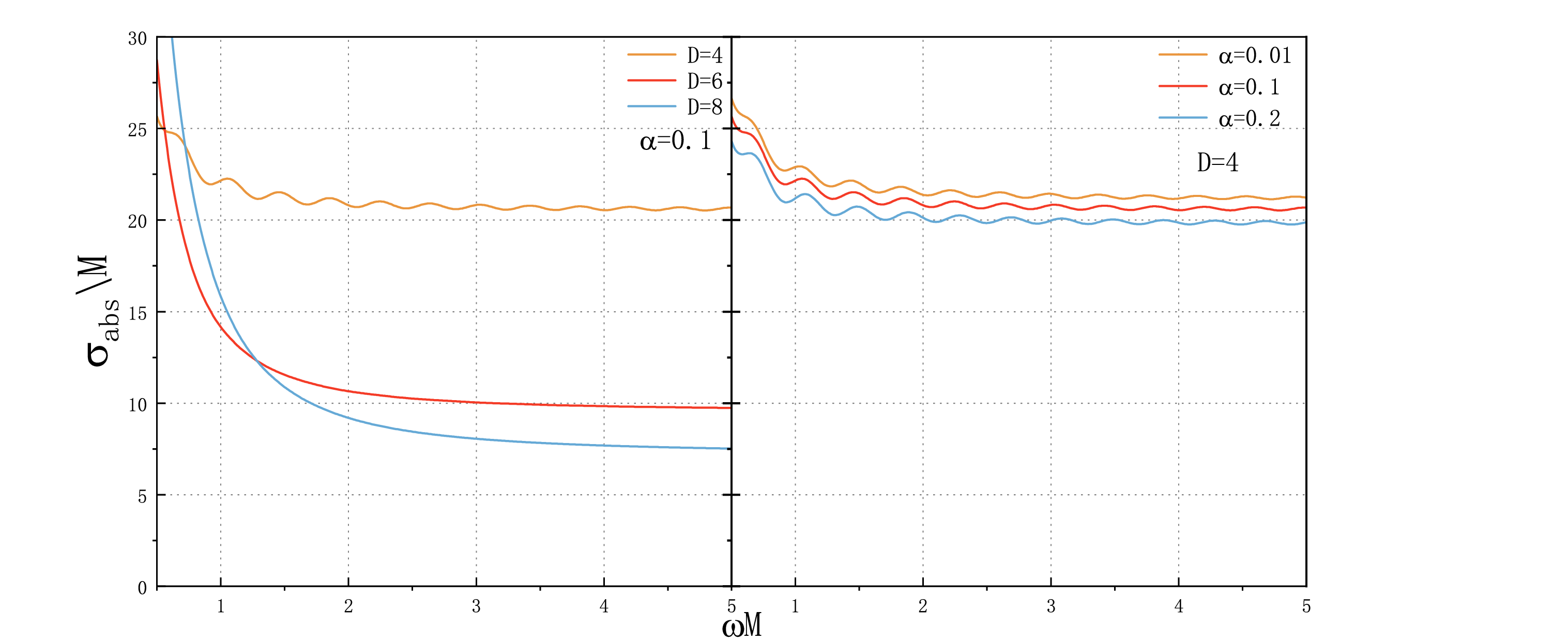}
     	\caption{\label{fig4} Left Figure illustrates the total absorption cross-section $\sigma_{\text{abs}}$ corresponding to the metric (\ref{eq4}) for different spacetime dimensions. Right Figure shows the total absorption cross-section $\sigma_{\text{abs}}$ corresponding to the metric (\ref{eq4}).}
     \end{figure}

    \begin{figure*}
    	\centering\noindent\hfil\includegraphics[scale=0.21]{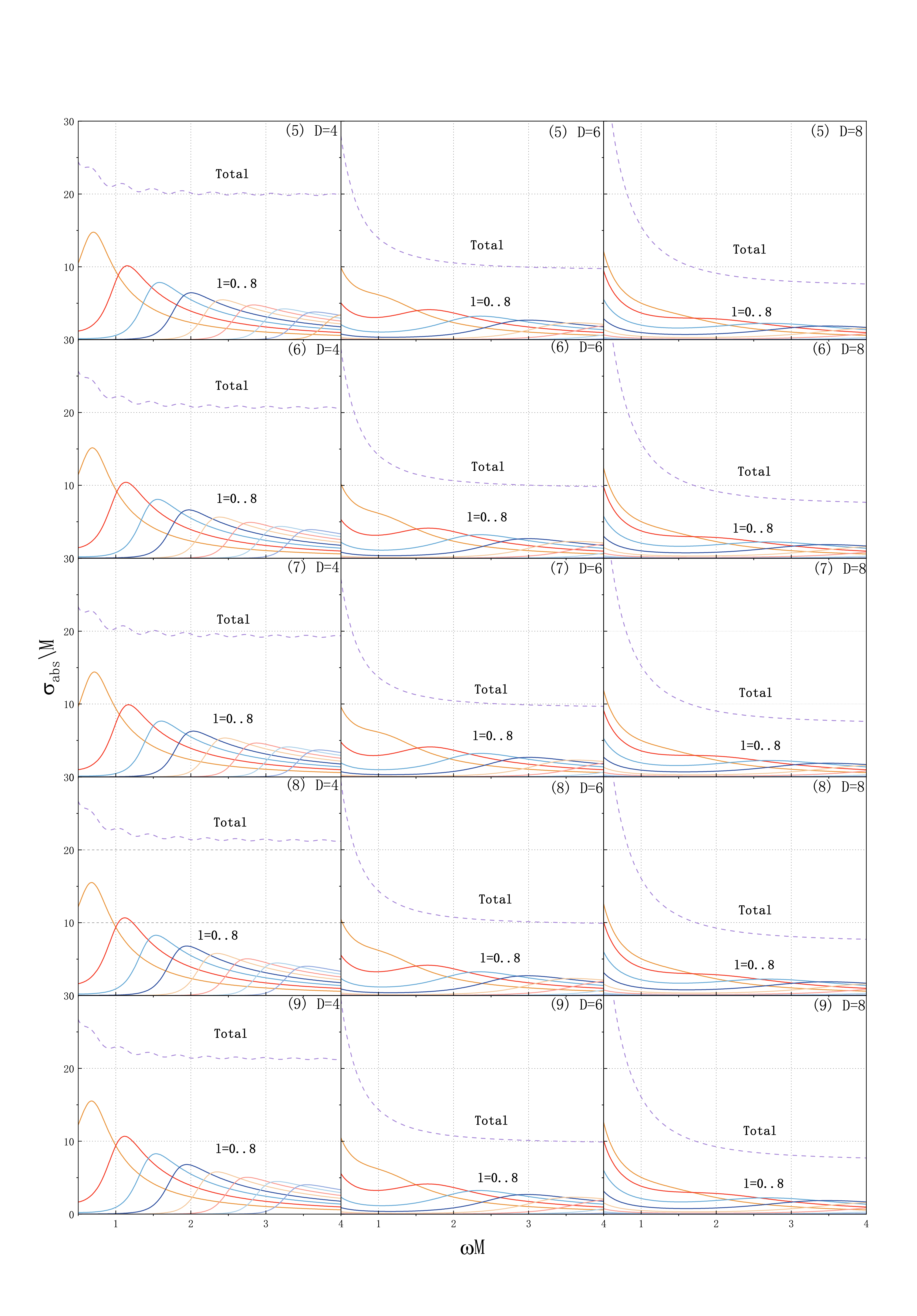}
    	\caption{
    		\label{fig5}Analysis of Partial and Total Absorption Cross-Sections for Different Black Hole Solutions.
    		}
    \end{figure*}

     Fig.~\ref{fig4} illustrates the behavior of the total absorption cross-section under fixed parameters $D$ and $\alpha$. Left figure depicts the total absorption cross-section $\sigma_{\text{abs}}$ as a function of spacetime dimension $D$ with fixed $\alpha$.
      The results indicate that Near the high-frequency limit ($\omega M \gg 1$), $\sigma_{\text{abs}}$ decreases with increasing $D$. In the low-frequency limit ($\omega M \ll 1$), the trend reverses. Further analysis of partial absorption cross-sections will be discussed later. Right figure shows the impact of parameter $\alpha$ on $\sigma_{\text{abs}}$ with fixed $D$. A clear decreasing trend in $\sigma_{\text{abs}}$ is observed as $\alpha$ increases.
     
     Fig.~\ref{fig5} displays the partial absorption cross-sections $\sigma_{\text{abs}}^{(l)}$ and the total absorption cross-section $\sigma_{\text{abs}}$ under varying $D$ and $\alpha$, with mass normalized to unity($M = 1$). Notably, $\sigma_{\text{abs}}^{(l)}$ gradually decreases with increasing $\omega M$. In the low-frequency limit, the $l = 0$ partial wave dominates the total absorption. For $D = 4$,$\sigma_{\text{abs}}^{(l)}$ exhibits regular oscillatory patterns for $l \geq 1$. As $D$ increases, oscillations at $l = 1$ weaken, and the onset of oscillations shifts to higher $l$.  
     In addition, the following conclusions can be drawn: First , The peak values of  $\sigma_{\text{abs}}^{(l)}$ diminish with increasing $l$ . Furthermore , Larger $l$ corresponds to higher $\omega M$ at peak positions, consistent with the rise in effective potential energy for higher $l$. This implies higher-energy partial waves are less absorbed by the potential barrier.  
     
     When $D \geq 4$, both $\sigma_{\text{abs}}^{(l)}$ and $\sigma_{\text{abs}}$ exhibit a monotonic decrease with increasing $D$.At $D = 4$, distinct peaks in $\sigma_{\text{abs}}^{(l)}$ are visible for different $l$. For $D \geq 5$, peak sharpness diminishes, which aligns with the observation that oscillations in $\sigma_{\text{abs}}$ become progressively weaker. Moreover, as the dimensionality increases beyond five, these oscillations in $\sigma_{\text{abs}}$ nearly vanish entirely.
     Furthermore,In the High-frequency limit, $\sigma_{\text{abs}}$ decreases with $D$.In the Low-frequency limit, Divergent behavior in $\sigma_{\text{abs}}^{(l)}$ leads to opposite trends in $\sigma_{\text{abs}}$ compared to high frequencies.

     \subsection{Scattering Cross-section}
     
     The expression for the scattering amplitude can be written as:  
     \begin{equation}  
     	g(\theta) = \sum_{l=0}^\infty (2l + 1) \left( e^{2i\delta_l} - 1 \right) P_l(\cos\theta),  
     \end{equation}  
     where $ P_l(\cos\theta) $ is the Legendre polynomial, and the differential scattering cross-section is defined as:  
     \begin{equation}  
     	\frac{d\sigma_{\text{sca}}}{d\Omega} = \left| g(\theta) \right|^2,  
     \end{equation}  
     From this, the scattering cross-section can be obtained:  
     \begin{equation}  
     	\sigma_{\text{sca}}(\omega) = \int \frac{d\sigma}{d\Omega} \, d\Omega = \frac{1}{2i\omega} \sum_{l=0}^\infty (2l + 1) \left| e^{2i\delta_l} - 1 \right|^2.  
     \end{equation}

      \begin{figure}
     	\noindent\hfil\includegraphics[scale=0.18]{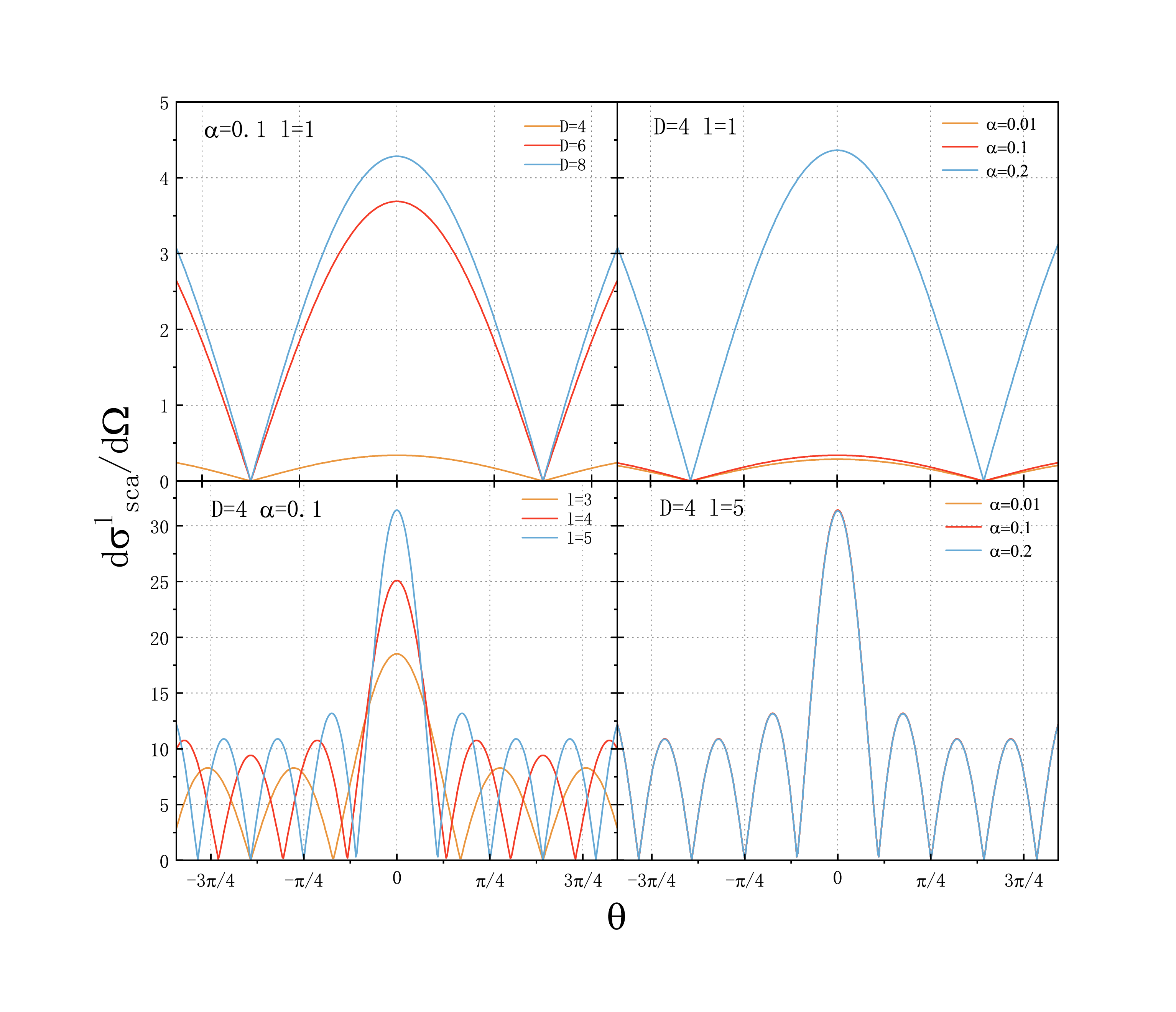}
     	\caption{\label{fig6}Figure demonstrates the differential scattering cross-section of Eqs.~(\ref{eq3}) under different parameters $\alpha$, $D$  and $l$.}
     \end{figure}

     Fig.~\ref{fig6} demonstrates the differential scattering cross section in the spacetime described by Eqs.~(\ref{eq3}). The following observations can be made:  
     Firstly, when parameters $\alpha$ and $D$ are fixed, the primary scattering angular width broadens with increasing $l$ \cite{Matzner1985}.When parameters $D$ and $l$ are fixed, the differential scattering cross-section increases with larger values of $\alpha$. This conclusion corresponds to our previous finding that the absorption cross-section decreases with increasing $\alpha$.
     Additionally, the differential scattering cross-section reaches its maximum value at $\theta = 0$.
     Lastly, when $l$ is large, the parameter $\alpha$ has a smaller impact on the differential scattering cross-section in the spacetime described by Eqs.~(\ref{eq4}).  
     
     In other words, the parameter $\alpha$ has a greater influence on low-frequency waves, a conclusion that has also been verified in the discussion of absorption cross-sections.

	\section{Conclusion}
     We investigate the scattering and absorption cross-sections of massless scalar fields for a family of spherically symmetric regular black hole solutions derived from quasitopological gravity. The interplay between the frequency $\omega M$ and parameters $\alpha$, $D$ is systematically studied, and provide numerical results for diverse parameter configurations.
     
     First, we analyze the QNMs of these black holes using the WKB approximation. Key findings are as follows:
     Both the real and imaginary parts of QNMs increase significantly with spacetime dimension $D$, leading to faster field decay in higher-dimensional black holes.
     Metric behavior differences are amplified at larger $D$, offering deeper insights into black hole characteristics.

     Next, we examine the absorption and scattering of massless scalar fields by these black holes. For $\alpha = 0.1$,the absorption cross-section $\sigma_{\text{abs}}$ decreases with increasing $D$.For $D=4$, $\sigma_{\text{abs}}$ exhibits oscillatory behavior near the high frequency geometric optics limit.In dimensions $D \geq 5$, $\sigma_{\text{abs}}$ monotonically decreases at high frequencies with suppressed oscillation amplitudes. A counterintuitive low-frequency enhancement of $\sigma_{\text{abs}}$ is observed, potentially linked to unique internal structures of the regular black hole solutions.

     Regarding the scattering cross-sections, we observe that with fixed $\alpha$ and $D$, the primary scattering angular width broadens progressively with increasing angular momentum quantum number $l$ \cite{Matzner1985}. Furthermore, our analysis reveals that enhancing the parameter $\alpha$ reduces the absorption cross-section $\sigma_{\mathrm{abs}}$ while simultaneously enhancing the scattering cross-section $\sigma_{\mathrm{sca}}$. This behavior demonstrates that $\alpha$ exerts a more pronounced influence on low-frequency wave interactions, a conclusion that aligns with our previous findings on absorption cross-section dependencies.
     
     {Recently, Aminov et al. \cite{Aminov2022} proposed an analytical approach based on N=2 supersymmetric gauge theory and quantum Seiberg-Witten curves, successfully mapping the spectral problem of black hole quasinormal modes (QNMs) to quantum period computations in gauge theory. By non-perturbatively reconstructing the divergent series in traditional WKB approximations through the Nekrasov-Shatashvili (NS) partition function, this method provides a convergent analytical framework for black hole spectral analysis. Notably, within five-dimensional gravitational theory, regular "Topological Star" solutions have been proposed as singularity-free alternatives to classical singular black holes\cite{Bah2021a,Bah2021b}. This approach avoids horizons and singularities through topological constraints and supersymmetric stabilization mechanisms, exhibiting equivalent macroscopic properties to the regular black hole solutions studied in this work while offering superior tractability under quantum field theory treatments due to their singularity-free microstructure. The connection between these approaches reveals a fundamental correspondence between higher-dimensional gravitational solutions and gauge theories, paving new pathways toward a unified description of quantum black hole behavior.}

    \bibliographystyle{unsrt}
	\bibliography{ref} 
	
\end{document}